\documentclass[12pt,preprint]{aastex}

\def\be{\begin{equation}}
\def\ee{\end{equation}}

\def\ergs{{\rm\,erg\,s^{-1}}}
\def\msun{M_{\odot}}

\def\ergs{\rm \,erg\,s^{-1}}
\def\be{\begin{equation}}
\def\ee{\end{equation}}
\catcode`\@=11 
\def\@versim#1#2{\vcenter{\offinterlineskip
        \ialign{$\m@th#1\hfil##\hfil$\crcr#2\crcr\sim\crcr } }}
\def\lsim{\mathrel{\mathpalette\@versim<}}
\def\gsim{\mathrel{\mathpalette\@versim>}}

\shorttitle{The nature of XBONGs}
\shortauthors{Feng Yuan \& Ramesh Narayan}
\begin{document}

\title{On the Nature of X-ray Bright Optically Normal Galaxies}
\author{Feng Yuan\altaffilmark{1,2} and
Ramesh Narayan\altaffilmark{2}}

\altaffiltext{1}{Department of Physics, Purdue University, West Lafayette,
IN 47907; fyuan@physics.purdue.edu}
\altaffiltext{2}{Harvard-Smithsonian Center for Astrophysics, 60 Garden Street,
Cambridge, MA 02138; narayan@cfa.harvard.edu}

\begin{abstract}

Recent X-ray surveys by {\it Chandra} and {\it XMM-Newton} have
revealed a population of X-ray bright, optically normal galaxies
(XBONGs) at moderate redshifts.  We propose that many XBONGs are
powered by an inner radiatively inefficient accretion flow (RIAF) plus
an outer radiatively efficient thin accretion disk.  The absence of
optical/UV activity in XBONGs is explained by the truncation of the
thin disk near the black hole, while the relatively strong X-ray
emission is explained as inverse Compton emission from the hot RIAF.
As an example, we show that the spectra of two XBONGs can be fit
fairly well with such a model.  By comparing these two sources to
other accreting black holes, we argue that XBONGs are intermediate in
their characteristics between distant luminous active galactic nuclei
and nearby low-luminosity nuclei.

\end{abstract}



\section{Introduction}

Recent deep X-ray surveys with {\it ROSAT, Chandra} and {\it
XMM-Newton} have resolved more than 80\% of the 0.1-10 keV cosmic
X-ray background into discrete sources (Hasinger et al. 1998;
Mushotzky et al. 2000; Giacconi et al. 2001; Brandt et al. 2001).
Optical identifications of these X-ray sources show that many are, as
expected, luminous active galactic nuclei (AGNs).  However, rather
unexpectedly, a sizeable number of relatively bright X-ray sources
have been spectroscopically identified with early-type ``normal''
galaxies without any obvious signature of nuclear activity in their
optical spectra (Fiore et al. 2000; Hornschemeier et al. 2001;
Giacconi et al. 2001; Barger et al. 2001; Comastri et al. 2002a,b).

The existence of this unusual class of sources was already pointed out
by Elvis et al. (1981) more than 20 years ago, based on an analysis of
{\it Einstein} observations.  It was later confirmed by Griffiths et
al. (1995) using {\it ROSAT} data.  The sources have been given a
variety of names, such as {\it Optically Dull Galaxies} (Elvis et
al. 1981), {\it Passive Galaxies} (Griffiths et al. 1995), {\it X-ray
Bright Optically Normal Galaxies} (XBONGs, Comastri et al. 2002b), and
{\it Elusive Active Galactic Nuclei} (Maiolino et al. 2003). We will
adopt the name XBONG in this paper.  According to Maiolino et
al. (2003), the fraction of XBONGs among local galaxies is comparable
to or even higher than that of optically selected Seyfert nuclei.

The large X--ray--to--optical flux ratios of XBONGs, as well as their
hard spectra in X--rays (at least in the brighter sources for which
spectral analysis is possible), suggest that AGN activity is
occurring in these objects.  The lack of evident optical emission
lines is, however, a puzzle.  There are several possible explanations.

One explanation is that XBONGs are luminous AGNs that happen to be
heavily obscured.  The obscuration must be in all directions, not just
in a torus, since the sources lack both broad lines and narrow lines
in their spectra (Marconi et al. 1994, 2000; Genzel et al. 1998; Spoon
et al. 2000; Fabian 2003).  Dudley \& Wynn-Williams (1997) predicted
that a deep silicate absorption at $9.7 \mu{\rm m}$ should be detected
in this case.  The feature has been seen in two XBONGs (NGC~4945:
Maiolino et al. 2000; NGC~4418: Spoon et al. 2001), indicating that
the complete obscuration idea is correct for at least some XBONGs.

Despite this success, it seems unlikely that the obscuration model
applies to all XBONGs.  Severgnini et al. (2003) performed a detailed
spectral analysis of three XBONGs observed with {\it XMM-Newton} and
found that only one out of the three is X-ray obscured ($N_H \approx
2\times 10^{23}~ {\rm cm}^{-2}$), while the other two sources are
relatively unobscured ($N_H \approx 4\times10^{21}, ~1\times10^{21} ~{\rm
cm^{-2}}$, respectively).  A similar result was obtained by Page et
al. (2003), who carried out optical spectroscopy of a number of X-ray
sources from the 13~hr {\em XMM-Newton/Chandra} deep survey.  Of their
70 sources, 23 were found to be XBONGs, half of which were unabsorbed
in X-rays.

A second explanation for XBONGs is that proposed by Moran, Filippenko,
\& Chornock (2002) and Severgnini et al. (2003) according to which the
lack of significant emission lines in XBONGs may be due to dilution of
the nuclear spectrum by starlight from the host galaxy. However, the
intrinsic optical continuum emission from XBONGs is weaker than in
luminous, quasar-like AGNs (e.g., Comastri et al 2002b), which
suggests that the emission-line luminosities are also intrinsically
low.  In terms of the spectral index $\alpha_{ox}$ (defined by
$F_{\nu} \propto \nu ^{-\alpha_{\rm ox}}$) between 2500~\AA~and 2 keV,
Severgnini et al. (2003) find that $\alpha_{ox} \approx 1.2$ for a
small sample of XBONGs. This value is systematically smaller than
$\alpha_{ox}\approx 1.5$ for luminous AGNs (Brandt, Laor \& Wills
2000).

A third possibility is that XBONGs are BL Lac-like
objects. Observations by {\it Chandra} and {\it XMM-Newton} of at
least one XBONG, the source ``P3'' (\S2.2), indicate no significant
flux or spectral variability over a time interval of seven months
(Comastri et al. 2002a).  This is highly unusual for a BL Lac object.
The presence of a large calcium break and a lack of detectable radio
emission in several XBONGs (Fiore et al. 2000) are additional
arguments against the BL Lac model.  Nevertheless, this model is hard
to rule out and may well describe some XBONGs, though it seems
unlikely to apply to a majority of XBONGs.

The last possibility, the one we focus on in this paper, is that the
intrinsic weakness of XBONGs in the optical/UV waveband is because
these sources lack an optically-thick accretion disk at small radii.
Instead of a cool disk, we suggest that the gas at small radii is in
the form of a very hot radiatively inefficient accretion flow (RIAF,
also known as an advection-dominated accretion flow or ADAF).  We
discuss this model in \S 2 and show that it is qualitatively
consistent with the spectra of two XBONGs (\S\S 2.1, 2.2).  We
conclude in \S 3 with a summary and discussion.

\section{RIAF Model of XBONGs}

The redshift and luminosity distribution of XBONGs provide clues to
the nature of these sources.  In a sample of 71+45 hard X--ray sources
selected from the {\tt HELLAS2XMM} and {\it Chandra} surveys, Comastri
et al. (2002b) found 10 XBONGs.  The 10 sources are at low redshift
and have low luminosities.  Brusa et al. (2003) have presented optical
identifications for a sample of 35 sources, selected again from the
{\tt HELLAS2XMM} survey.  They find that at low redshift, the sources
consist of a mix of Broad Line AGNs, Narrow Emission-Line galaxies and
XBONGs, whereas at $z> 1$ all the sources are spectroscopically
classified as Broad Line AGNs.  Similarly, the 23 XBONGs identified by
Page et al. (2003) all lie at $z<1$ and have lower X-ray luminosities,
whereas the non-XBONG sources in their sample extend over a wide range
of redshift and are typically brighter.

These results indicate that XBONGs have lower luminosities compared to
standard AGNs.  The fact that the sources are also found
preferentially at lower redshift follows naturally.  Supermassive
black holes in the nuclei of high redshift galaxies are generally
believed to have a lot of gas available for accretion and are
therefore quite luminous.  At lower redshifts there should be less gas
and we expect the nuclei to be less luminous.  Therefore, if XBONG
behavior is associated with lower luminosities, then the sources
should be found at somewhat lower redshifts than bright AGNs.  But why
should lower luminosity AGNs have proportionately less optical
emission compared to more luminous objects?  We propose that these
sources do not have a standard geometrically-thin optically-thick disk
at small radii.  Instead, the disk is truncated at a ``transition''
radius $R_{\rm tr}$, and the accretion flow inside $R_{\rm tr}$ is in
the form of an optically-thin RIAF.  (Early versions of the RIAF model
were called ADAFs, Narayan \& Yi 1995; see Narayan, Mahadevan \&
Quataert 1998 and Kato, Fukue, \& Mineshige 1998 for reviews).
Because of the absence of an optically-thick disk at small radii, the
source emits much less optical and UV radiation.

This paradigm --- an inner RIAF plus an outer standard thin disk ---
has been shown to work very well (even required) in several types of
sources: black hole X-ray binaries in the hard state (Esin, McClintock
\& Narayan 1997; Poutanen, Krolik \& Ryde 1997; Dove et al. 1998; see
McClintock \& Remillard 2003 and Zdziarski \& Gierlinski 2004 for
reviews), low-luminosity AGNs in our local universe (Quataert et
al. 1999; Shields et al. 2000), and some Seyfert 1 galaxies (Chiang
2002; Chiang \& Blaes 2003).  Quataert et al. (1999) estimated $R_{\rm
tr}\approx 100R_s$ for the LLAGNs, M~81 and NGC~4579, while Chiang \&
Blaes (2003) estimated $R_{\rm tr}\approx 30-80R_s$ for the Seyfert 1
galaxy NGC~5548.  The lack of a big blue bump in the spectra of LLAGNs
(Ho 1999) indicates that an optically-thick disk is not present at
small radii in these sources, and there is a similar suggestion in the
case of NGC~5548 from the lack of a relativistically broadened and
skewed Fe K$\alpha$ line in its X-ray spectrum (Pounds et al. 2003).
As in the case of the XBONGs, all these sources have relatively low
optical luminosities compared to quasars.  M81 and NGC~4579 have
bolometric luminosities $\approx 10^{-4}$ Eddington, while NGC~5548 has a
luminosity of $\approx 10^{-2}$ Eddington.  Finally, as mentioned in \S
1, Severgnini et al. (2003) measured a spectral index $\alpha_{\rm ox}
\approx 1.2$ for three XBONGs, which is lower than $\alpha_{\rm ox}\approx
1.5$ for typical AGNs.  In comparison, LLAGNs have $\alpha_{\rm ox}
\approx 0.75$---$1.08$, with a mean value of $0.9$ (Ho 1999), and
NGC~5548 has $\alpha_{\rm ox} \approx 1.1$ (Chiang \& Blaes 2003).

In view of all these indications that XBONGs are spectrally similar to
LLAGNs and some Seyfert 1 galaxies, it is reasonable to suppose that
the same accretion geometry, viz., an inner RIAF plus an outer thin
disk, is present in all the sources.  Comastri et al. (2002a) have
ruled against such a model for XBONGs using two different arguments.
First, they used the variability trend found by Nandra et al. (1997)
to argue that if XBONGs are low-luminosity AGNs, they should be highly
variable, whereas there is no evidence for significant variability.
However, Ptak et al. (1998) have shown that AGNs with luminosities
below a certain threshold do not follow the Nandra et al. trend.
Indeed, Ptak et al. cite this as evidence that LLAGNs have
ADAFs/RIAFs.  Second, Comastri et al. (2002a) compared the spectrum of
the XBONG source P3 with certain model spectra from Quataert \&
Narayan (1999) and claimed that the radio flux of the source is
incompatible with a RIAF model. However, such a comparison is very
crude; for example, the effect of the mass of the black hole was not
taken into account. In fact, the detailed calculations presented below
show that a RIAF model is consistent with the data.

We now analyze in detail two XBONG sources and show that the model
explains the main features of their spectral energy distributions.

\subsection{Source \#1}

The spectral data available on XBONGs is generally rather limited,
with XMMJ021822.3-050615.7, or ``Source \#1'' (Severgnini et
al. 2003), having perhaps the best data (see Figure 1).  A radio flux
is available from {\it VLA} observations, and X-ray data have been
obtained with {\it XMM-Newton}.  The X-ray luminosity is $L_{\rm
2-10keV}=5.6 \times 10^{42}\ergs$, and the X-ray photon index is
$\Gamma=1.66\pm0.30$.  Severgnini et al. (2003) measured $\alpha_{ox}
=1.2\pm0.1$, which gives the 2500{\AA} data point shown in Figure 1.
Using the Magorrian et al. (1998) relation between black hole mass and
bulge luminosity, we estimate the black hole mass to be $M_{\rm bh}
\approx 3 \times 10^9\msun$.  This value should be regarded as an
upper limit (Ferrarese \& Merritt 2000).

We have modeled the spectrum of Source \#1 using a current version of
the RIAF model.  Recent numerical simulations (Stone, Pringle, \&
Begelman 1999; Hawley \& Balbus 2002; Igumenshchev et al. 2003) and
analytical work (Blandford \& Begelman 1999; Narayan et al. 2000;
Quataert \& Gruzinov 2000) indicate that probably only a fraction of
the gas that is available at large radius in a RIAF actually accretes
onto the black hole. The rest of the gas is either ejected from the
flow or is prevented from accreting by convective motions.  However,
the details are likely to depend on the accretion rate.  In the case
of XBONGs, since the accretion rate is probably moderately high, the
Bernoulli parameter as well as the entropy gradient are likely to
approach zero.  As a result, outflows and convection may be less
well-developed.  Given the theoretical uncertainty, we consider two
different kinds of models. In the first model, we assume that the
accretion rate in the RIAF varies with radius as
\begin{equation}
\dot{M}_{\rm RIAF}(R)= \dot{M}_0 \left({R\over R_{\rm tr}}\right)^s,
\qquad R \le R_{\rm tr}.
\end{equation}
Here $R_{\rm tr}$ is the transition radius between the outer thin disk
and the inner RIAF, and $\dot{M}_0$ is the accretion rate of the RIAF
at $R_{\rm tr}$.  We assume $s=0.3$ as in the case of Sgr A* (Yuan,
Quataert \& Narayan 2003), and we correspondingly set $\delta$, the
fraction of the turbulent energy that heats the electrons, equal to
0.5 (see Quataert \& Gruzinov 1999). In the second model, we assume
that the accretion rate is independent of radius, i.e., $s=0$, and we
correspondingly set $\delta=10^{-2}$ (as in the original ADAF model,
e.g., Narayan et al. 1998).  We solve the radiation-hydrodynamic
equations of the RIAF self-consistently to obtain the profiles of
density and electron temperature, and we then calculate the emergent
spectrum (see Yuan, Quataert \& Narayan 2003 for details).  The
radiative processes we consider in the RIAF include synchrotron and
bremsstrahlung emission, and the Comptonization of these photons as
well as soft photons from the outer thin disk.  For the thin disk, we
assume that the emitted spectrum is locally a blackbody.  We include
both the energy from viscous dissipation within the disk and the
reprocessing of hot radiation impinging from the RIAF.

The thick solid line in Figure 1 shows the predicted spectrum of the
first model (with $s=0.3$, $\delta=0.5$) for $\dot{M}_0=
10^{-2}\dot{M}_{\rm Edd}$ and $R_{\rm tr}=60 R_s$.  The dashed line
shows the spectrum from the RIAF alone and the dot-dashed line from
the outer thin disk alone.  The combined spectrum fits the optical and
X-ray data quite well. The under-prediction of the radio data is
common in RIAF models (e.g., Quataert et al. 1999). The flux would be
enhanced if the source has a jet (e.g., Yuan, Markoff, \& Falcke
2002a; Yuan et al. 2002b), or if the RIAF has nonthermal electrons (e.g., Mahadevan
1999; \"Ozel, Psaltis \& Narayan 2000; Yuan, Quataert \& Narayan 2003).  We have not
taken these effects into account in the present paper.  The transition
radius in the model is a little smaller than in the models of M81 and
NGC 4579 ($\approx 100R_s$; Quataert et al. 1999), but appear to be
similar to that found for the Seyfert 1 galaxy NGC~5548 ($\approx
30-80 R_s$; Chiang \& Blaes 2003).  The smaller radius explains why
$\alpha_{\rm ox}$ is larger in Source \#1 and NGC~5548 compared to
LLAGNs. The thin solid line in Figure 1 shows the result of the second
model (with $s=0, \delta=10^{-2}$). Because of a degeneracy between $s$ and  
$\delta$ (see Quataert \& Narayan 1999), the parameters are very similar
to the thick solid line, $\dot{M}=10^{-2}\dot{M}_{\rm Edd}$ and 
$R_{\rm tr}=40 R_s$.

The low optical flux of the model is a direct consequence of
truncating the thin disk at $R_{\rm tr}$.  For comparison, we show by
the three dotted lines the emission from three standard thin disks
extending down to the innermost stable circular orbit (ISCO) at
$3R_S$, with accretion rates of $\dot{M}_0= 10^{-3}, ~2\times
10^{-4}$, and $5 \times 10^{-5} \dot{M}_{\rm Edd}$, respectively.  The
upper two models predict too much flux in the optical and are
immediately ruled out.  Although the model with $\dot{M}_0=5\times
10^{-5} \dot{M}_{\rm Edd}$ can reproduce the 2500~\AA~ flux, it would
require the X-ray flux (which is presumably produced by a corona above
the disk) to be $ \approx 50\%$ of the bolometric luminosity, which is
highly unusual for an AGN.

In the above models, the mass of the black hole we have adopted is an
upper limit. It is not easy to estimate the uncertainty in the mass,
but fortunately this parameter does not affect any of the conclusions.
For instance, if the black hole mass is 10 times lower, i.e., $M_{\rm bh}
\approx 3 \times 10^8\msun$, we only need
to change the parameters to $\dot{M}_0\approx 2 \times
10^{-2}\dot{M}_{\rm Edd}$ and $R_{\rm tr}\approx 55 R_s$, while keeping the
other parameters unchanged ($s=0.3, \delta=0.5$), 
to match the data.  Also, we find
that a standard thin disk extending down to the ISCO is still
inconsistent with the data.

\subsection{Source P3}

The second object we analyze is the source P3, which is the first
known example of an XBONG (Fiore et al. 2000).  This source is
interesting because Comastri et al. (2002a) claimed that the RIAF
model cannot fit its spectrum.  The mass of the central black hole is
estimated from the observed B luminosity of the galaxy bulge to be
$M_{\rm bh}\approx 4\times 10^8 \msun$ (Fiore et al. 2000).  As before,
this is an upper limit on the mass.

Multiwavelength observations of P3 were carried out by Comastri et
al. (2002a) and the results are plotted in Figure 2.  Radio
observations at 5 GHz give a 3 $\sigma$ upper limit of 0.15 mJy.
Near-infrared and optical observations again give only upper limits
which are not very useful to constrain models.  We show the optical
upper limits in Figure 2. \footnote{Note that an extinction-correction
has been applied to the optical limits using the $N_{\rm H}$ value
derived from {\it XMM-Newton} data} In addition, we arbitrarily assume
that the value of the spectral index $\alpha_{\rm ox}$ is identical to
that of Source \#1, i.e., $\alpha_{\rm ox}= 1.2$. The approximate
optical luminosity at 2500~\AA ~thus calculated is indicated by the
filled circle in the optical band.  P3 was observed in X-rays by both
{\it Chandra} and {\it XMM-Newton}.  Unfortunately, due to the
weakness of the source, only a small number of counts were obtained,
$\approx 60$ and $200$ in the two observations.  The {\it Chandra} data
give a photon index of $\Gamma \approx 1.4$, assuming a Galactic
column density.  For the {\it XMM-Newton} observations, acceptable
fits are obtained with a power-law of photon index $\Gamma=1.1\pm
0.35$, or $\Gamma=1.8$ (an average value for AGNs). The latter fitting
gives an absorption-corrected 2-10 keV luminosity of about $3\times
10^{42} \ergs$. The X-ray ``bow-tie'' corresponds to the two extremes
of the X-ray data, with two photon indices $\Gamma = 1.1$ and 1.8.

Comastri et al. (2002a) considered the obscuration model in detail for
P3.  Assuming an average value of the optical-to-X-ray flux ratio
typical of hard X-ray-selected unobscured quasars, and taking a
standard Galactic extinction curve, they calculated the expected
optical magnitude of the source corresponding to the best-fit $N_{\rm
H}$ from the X-ray analysis.  They concluded that the nuclear emission
would probably have been detected in the optical band, in conflict
with observations.  In fact, even when they assumed that the obscuring
material is Compton-thick ($N_H \ga 1.5 \times 10^{24} ~{\rm cm^{-2}}$
for which there is no observational evidence), by comparing the
spectra of P3 and NGC 6240, the prototype of a heavily obscured AGN,
they found that although they could explain the absence of optical
activity in P3, the radio and IR flux of the source are significantly
lower than that of NGC 6240.

The thick solid line in Figure 2 shows the spectrum predicted by a
RIAF model of P3 with $\dot{M}_0=1.3 \times 10^{-2}\dot{M}_{\rm Edd}$,
$R_{\rm tr}=60 R_s$. These parameters are similar to those of Source
\#1.  The dashed line shows the spectrum from the RIAF alone and the
dot-dashed line from the outer thin disk alone.  The combined spectrum
satisfies all the available constraints.  In particular, we see that
the model is consistent with the upper limit on the radio flux,
contrary to the claim of Comastri et al. (2002a).  The three dotted
lines show the emission from three standard thin disks extending down
to the ISCO, with accretion rates of $\dot{M}_0=8\times 10^{-5},
8\times 10^{-4}$, and $10^{-2} \dot{M}_{\rm Edd}$, respectively. As in
the case of Source \#1, none of these models fits the spectrum well.

\section{Summary and Discussion}

We have investigated in this paper the nature of the population of
X-ray bright, optically normal galaxies, or XBONGs.  While the absence
of optical activity in some XBONGs may be due to obscuration, this
explanation is unlikely to apply to all XBONGs.  We consider the
possibility that many XBONGs possess radiatively inefficient accretion
flows (RIAFs, formerly ADAFs).  According to this model, the accretion
flow occurs as a geometrically-thin optically-thick disk for radii
larger than a transition radius $R_{\rm tr}$, and as an optically-thin
RIAF for radii below $R_{\rm tr}$.  Because of the absence of an
optically-thick disk at small radii, there is very little optical or
UV emission, and therefore very little broad or narrow line emission.
Since RIAFs are possible only at small accretion rates (Narayan \& Yi
1995; Esin et al. 1997), XBONGs are expected to populate the
low-redshift, low-luminosity end of the AGN distribution. This is
confirmed by observations.

XBONGs are spectrally very similar to LLAGNs in our local universe as
well as some Seyfert 1 galaxies. All of these sources have weak
optical/UV emission and relatively large X-ray fluxes.  The spectral
index $\alpha_{ox}$ between 2500\AA~and 2keV is $\approx 1.2$ for
XBONGs, $\approx 1.1$ for the Seyfert 1 galaxy NGC~5548, and $\approx
0.9$ for a small sample of LLAGNs.  These values are small compared to
$\alpha_{ox}\approx 1.5$ for luminous AGNs.  Both LLAGNs and NGC~5548
have been successfully modeled by RIAFs (Quataert et al. 1999; Chiang
\& Blaes 2003).  The case for considering a similar model for XBONGs
is thus strong.  To test this proposal, we have modeled two XBONGs,
Source \#1 (\S 2.1) and P3 (\S 2.2). We find that the RIAF model fits
the available spectral data on both sources reasonably well (Figures
1, 2). We also find that any model in which the thin disk comes all
the way down to the innermost stable circular orbit (ISCO) is
inconsistent with the data.

It is useful to place XBONGs in the context of other accreting black
hole sources.  Figure 3 shows the transition radius $R_{\rm tr}$ in
Schwarzschild units and the luminosity $L$ in Eddington units for a
number of objects whose spectra have been fitted with the RIAF model.
Luminous black hole sources, including AGNs at high redshifts and
black hole X-ray binaries in the high soft state, have standard
radiatively efficient accretion disks extending down to the ISCO.
When the luminosity is below a certain value, say $\approx 0.03L_{\rm
Edd}$, an advection-dominated RIAF is allowed (Esin et al. 1998), and
it is postulated that the thin disk is then truncated and the
innermost region of the flow is replaced by a RIAF.  It is also
expected that the lower the luminosity $L/L_{\rm Edd}$ the larger the
transition radius $R_{\rm tr}/R_s$ (e.g., Narayan \& Yi 1995; Esin et
al. 1997; Narayan et al. 1998; R\'{o}za\'{n}ska \& Czerny 2000;
Manmoto \& Kato 2000).  Such a correlation is clearly seen in Figure
3.  In order of decreasing luminosity $L/L_{\rm Edd}$, the sources
shown are: 1) X-ray binaries in the soft state and bright AGNs; 2)
X-ray binaries in the hard state and Seyfert 1 galaxies; 3) XBONGs; 4)
LLAGNs; 5) X-ray binaries in the quiescent state; 6) ultra-dim AGNs.
Note that we have analyzed only two XBONGs in this paper, both of
which appear to be brighter than typical LLAGN when measured in
Eddington units.  As a class, however, XBONGs are quite heterogeneous,
and it is possible that other XBONGs are similar in luminosity to
LLAGNs, or perhaps even dimmer.

One caveat to note is that our estimates of $R_{\rm tr}/R_S$ for
Source \#1 and P3 are sensitive to the assumed optical fluxes of the
two sources.  The optical flux of Source \#1 was obtained by
Severgnini et al. (2003) by an indirect method, while P3 has no
optical measurement and we have arbitrarily assumed the same flux as
in Source \#1.  Our estimates of the transition radii are thus rather
uncertain.

Finally, while we have suggested in this paper that a RIAF is present
in many XBONGs, we do not mean to imply that the RIAF necessarily
produces the entire emission. It is possible, for instance, that some
of the radiation (e.g., radio) comes from a jet.  The BL Lac-type
model mentioned in \S 1 is an extreme example of this scenario in
which most of the emission comes from a highly relativistic and beamed
flow.  Another significant zone of emission is the base of the jet, or
the region where the jet and the disk meet (e.g., see Livio, Pringle
\& King 2003).  The Doppler factor here is expected to be small so
variability should be weak.  A model of this region has been
successfully worked out for Sgr A* (Yuan, Markoff, \& Falcke 2002a)
and NGC~4258 (Yuan et al. 2002b).

\begin{acknowledgements}
We thank our second referee for a number of useful comments which
helped improve the manuscript greatly.  This work was supported in
part by NSF grant AST-0307433 and NASA grants NAG5-9998 and
NAG5-10780.

\end{acknowledgements}

{} 

\clearpage

\begin{figure}
\epsscale{0.90}
\plotone{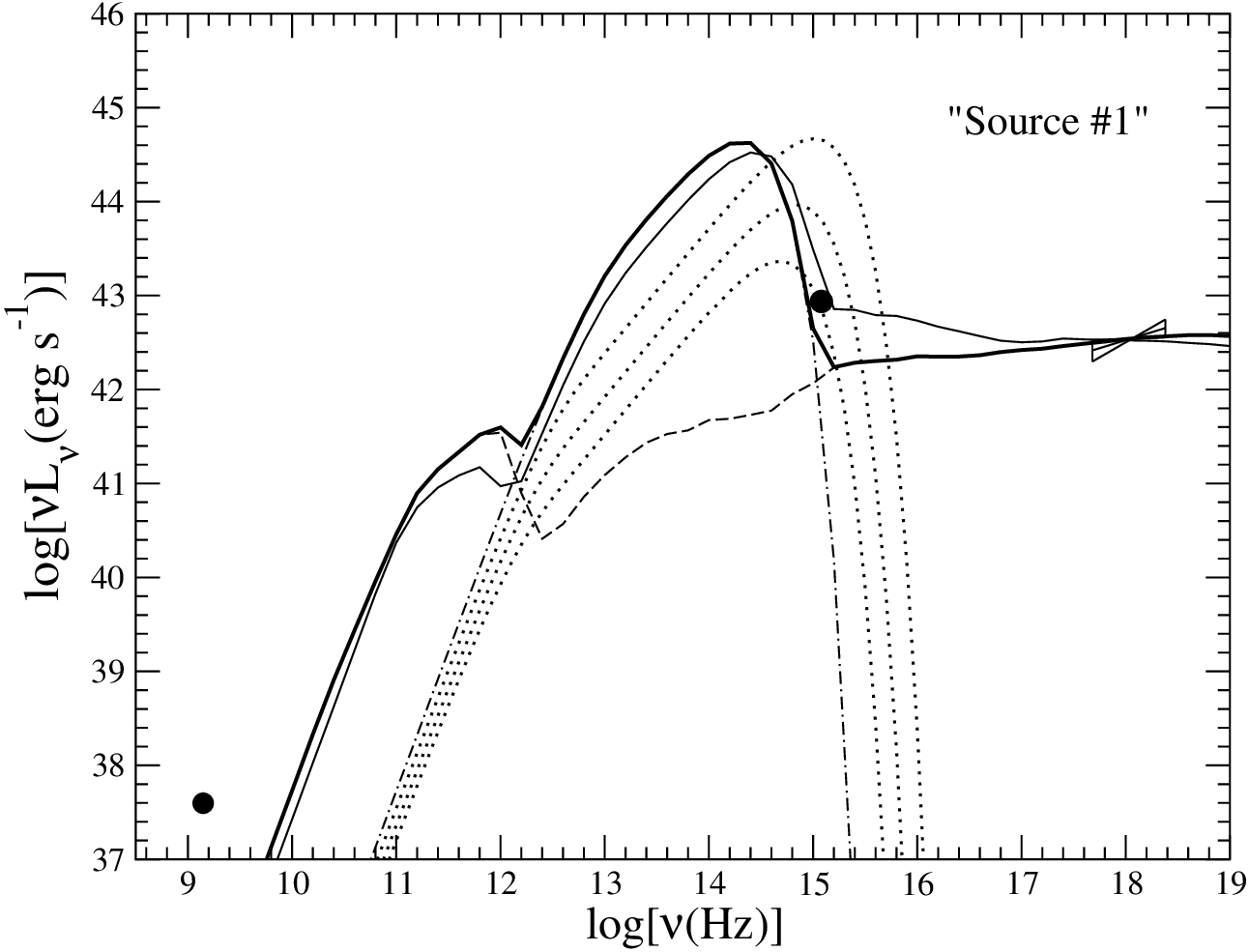}
\vspace{.0in}
\caption{Spectral fit of the XBONG ``Source \#1'' (Severginini et
al. 2003) with a RIAF+thin disk model. The thick solid line is the
combined spectrum predicted for an accretion flow consisting of a
truncated thin disk for radii $R>R_{\rm tr} = 60R_S$ (dot-dashed line)
and a RIAF for $R<R_{\rm tr}$ (dashed line). The mass accretion rate
of the RIAF at $R=R_{\rm tr}$ is $\dot{M}_0= 
10^{-2}\dot{M}_{\rm Edd}$ and it decreases with radius according to
equation (1) with $s=0.3$. The thin solid line shows the result of
another model (traditional ADAF) in which the accretion rate of the
RIAF is taken to be independent of radius, with $\dot{M}=
10^{-2}\dot{M}_{\rm Edd}$ and $R_{\rm tr} = 40R_S$. The three dotted lines
show the emission from three standard thin accretion disks extending
all the way down to $R=3R_S$ with (from bottom to top)
$\dot{M}/\dot{M}_{\rm Edd}=5\times 10^{-5}, ~2 \times 10^{-4}$, and
$10^{-3}$. These latter models do not fit the observations.}
\end{figure}

\begin{figure}
\epsscale{0.90}
\plotone{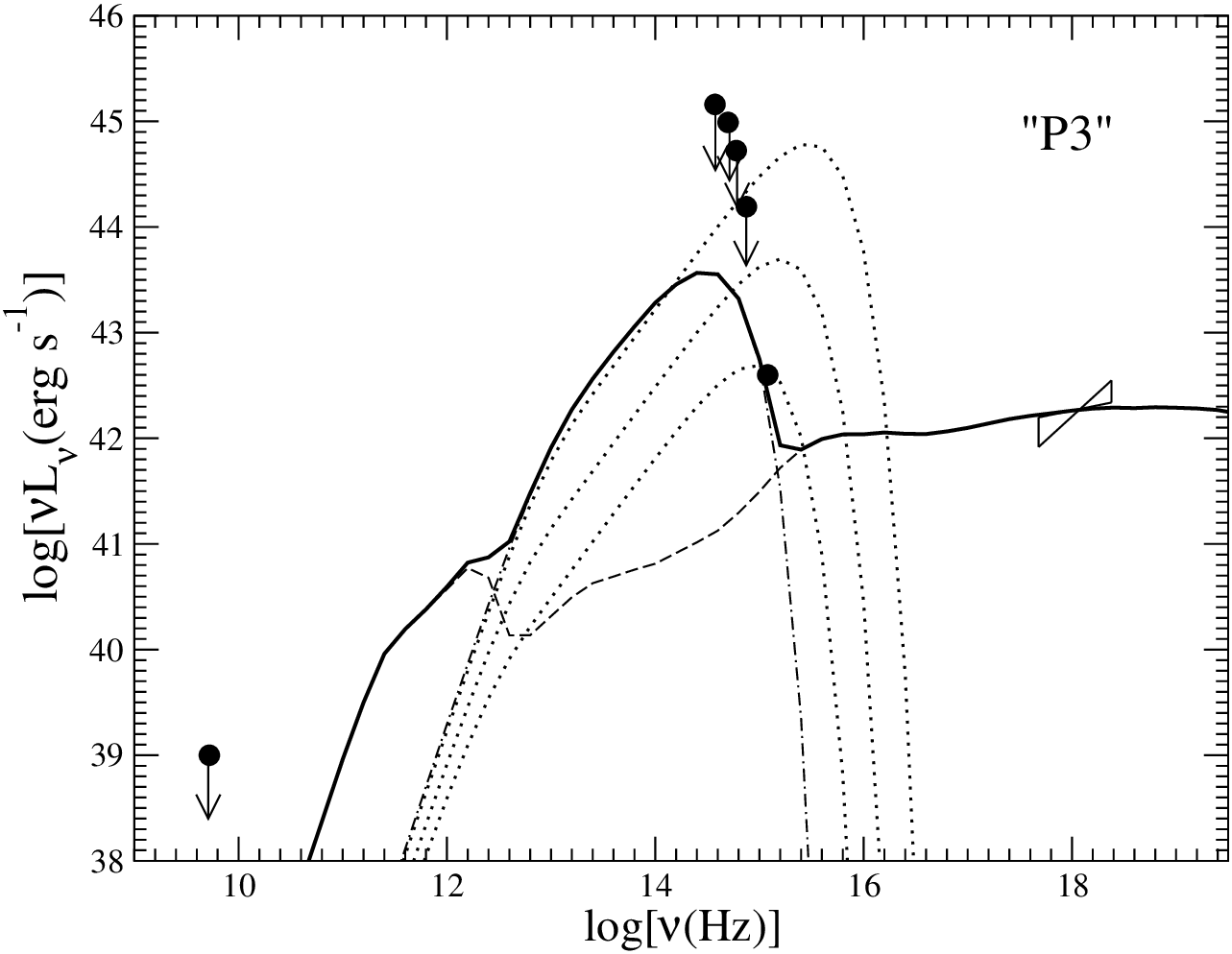}
\vspace{.2in}
\caption{Spectral fit of the XBONG source P3 with a RIAF+thin disk
model.  The thick solid line is the combined spectrum predicted for an
accretion flow consisting of a truncated thin disk for radii $R>R_{\rm
tr} = 60R_S$ (dot-dashed line) and a RIAF for $R<R_{\rm tr}$ (dashed
line). The mass accretion rate of the RIAF at $R=R_{\rm tr}$ is
$\dot{M}_0= 1.3\times 10^{-2}\dot{M}_{\rm Edd}$ and it decreases with
radius according to equation (1) with $s=0.3$.  The three dotted lines
show the emission from three standard thin accretion disks extending
all the way down to $R=3R_S$ with (from bottom to top)
$\dot{M}/\dot{M}_{\rm Edd}=8\times 10^{-5}, ~8 \times 10^{-4}$, and
$10^{-2}$.  These latter models do not fit the observations.}
\end{figure}

\begin{figure}
\epsscale{0.90}
\plotone{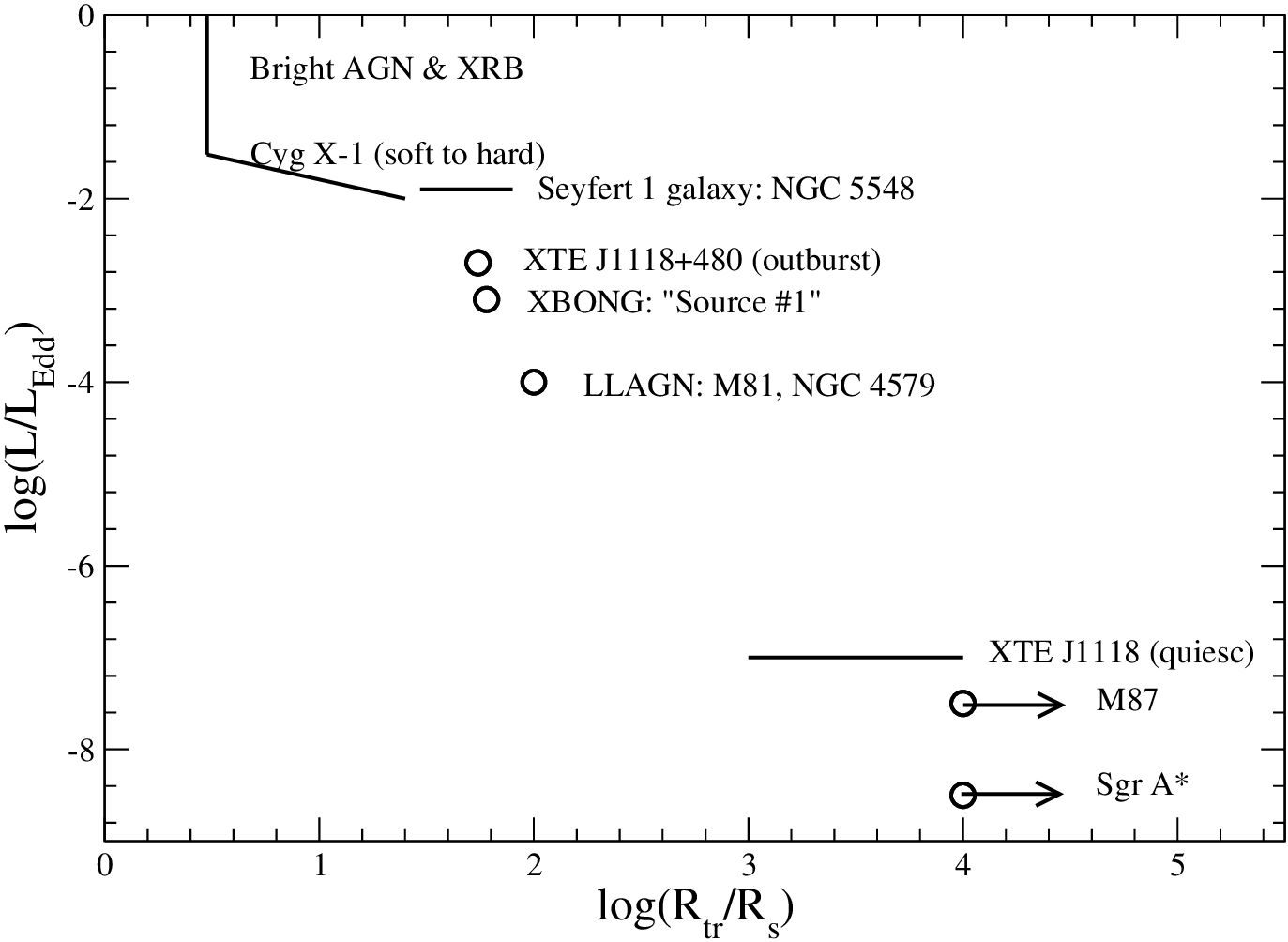}
\vspace{.3in}
\caption{Plot of the bolometric luminosity in Eddington units along
the ordinate versus the transition radius in Schwarzschild units along
the abscissa for different accreting black holes.  The slanting line
for Cyg X-1 is based on the model of Esin et al. (1998) for the
soft-to-hard state transition in this source.  The other data points
are from the following: Seyfert 1 galaxy NGC~5548 (Chiang \& Blaes
2003); outburst state of XTE J1118+480 (Esin et al. 2001); Source \#1
(this paper); M81 and NGC 4579 (Quataert et al. 1999); quiescent state
of XTE J1118+480 (McClintock et al. 2003); M~87 (Di Matteo et
al. 2003); Sgr A* (Yuan, Quataert \& Narayan 2003).}
\end{figure}

\end{document}